\documentclass[preprint, superscriptaddress, amssymb, floats,
byrevetex]{revtex4}
\usepackage{latexsym}
\usepackage{graphics}
\usepackage{amsmath,amsfonts,amssymb,amsthm,amstext,amscd,eucal}
\usepackage{bbold}
\usepackage{array}
\usepackage{hyperref}

\newcommand{\MUNCH}[1]{\relax}
\allowdisplaybreaks[1]
\numberwithin{equation}{section}
\hypersetup{%
  pdftitle = {Title},
  pdfkeywords = {keywords},
  pdfauthor  = {Joan Sim\'on}
}

\begin{document}
\title{
\normalsize \mbox{ }\hspace{\fill}\begin{minipage}{12 cm} {\tt
~~~~~~~~~~~~~~~~~~ UPR-1115-T, hep-th/0504136}{\hfill}
\end{minipage}\\[5ex]
{\large\bf Supersymmetric Kerr--Anti-de Sitter solutions
\\[1ex]}}
\date{\today}
\author{Mirjam Cveti{\v c}, Peng Gao and Joan Sim\'on}
\affiliation{ Department of Physics and Astronomy, University of Pennsylvania, \\
Philadelphia, PA 19104, USA}
\email{cvetic@cvetic.hep.upenn.edu; gaopeng@sas.upenn.edu;
jsimon@bokchoy.hep.upenn.edu}
\date{\today}

\begin{abstract}
We prove the existence of one quarter supersymmetric type IIB configurations
that arise as non-trivial scaling solutions of the standard five dimensional
Kerr-Anti-de Sitter black holes by the explicit construction of its Killing
spinors. This neutral, spinning solution is asymptotic to the static
anti-deSitter space-time with cosmological constant
$-\textstyle{\frac{1}{\ell^2}}$, it has two finite equal angular momenta
$J_1=\pm J_2$, mass $M=\textstyle{\frac{1}{\ell}}\,(|J_1|+|J_2|)$ and a naked
singularity. We also address the scaling limit associated with one half
supersymmetric solution with only one angular momentum.
\end{abstract}

\maketitle

\tableofcontents

\section{Introduction}

In the recent year important progress has been made in constructing general
R-charged, spinning solutions with  non-zero cosmological constant in
dimensions $D=5$ \cite{CHP,CHP2} and $D=7$ \cite{CCHP}. These solutions are
parameterized by a mass, three charges  in $D=5$ (two  charges in $D=7$) and
all angular momenta set equal. While the most general solutions with
all non-equal angular momenta are still elusive, these solutions
\cite{CHP,CHP2,CCHP} provide a  useful framework to address their
thermodynamics \cite{gary}, supersymmetric (BPS)  limits, such as
those found in $D=5$
\cite{KS,GR}, and their global space-time structure.  [For the earlier study
of the BPS limits of charged spinning solutions in $D=4$ see \cite{KP}.] In
particular, the BPS limits of these spinning charged solutions should play an
important role in elucidating the field configurations in the dual conformal
field theory. [For example, the singularity of the extremal limit of
the single R-charge spacetime \cite{myers} is interpreted as a
distribution of giant gravitons \cite{ST, myerssg, HHI}.]

On the other hand, the general  neutral (vanishing R-charge sector) spinning
solutions with cosmological constant in $D=5$ dimensions were constructed in
\cite{HHT} and subsequently in all dimensions $D>5$ in \cite{GLPP}. Their
thermodynamics has been studied extensively in \cite{GPP}. However,  the study
of their  BPS limits, except in $D=3$, seems to have led to  negative
conclusions \cite{HHT}.  On the other hand, there is an expectation that in
the dual field theory there should be field configurations that in the
strongly coupled limit represent BPS configurations with the  role of the spin
being associated with the angular momentum.
 The purpose of this paper is to prove explicitly the
  existence of supersymmetric,
asymptotically Anti-de Sitter (AdS)  space-times  with finite energy and
finite angular momentum, but no R-charge.

The existence of such BPS configurations is   not forbidden from a
supersymmetry algebra point of view. In the analysis done by \cite{KP,C1,C2}
in four dimensions, one can certainly saturate the Bogomolnyi type bound in
the zero R-charge sector. In principle similar conclusions can be derived in
five dimensions from the general analysis of the Bogomolnyi bounds presented
in \cite{gary}. The question remains, though, as to whether one can find
explicit classical configurations carrying these global charges, and to show
explicitly that these, Bogomolnyi bound saturating configurations, are indeed
supersymmetric.

While we believe the analysis of such BPS configurations   can be generalized
to non-equal angular momenta and  other dimensions, we shall focus on the case
in $D=5$ and two angular momenta equal. In the last section we shall also
discuss the case that corresponds to  the supersymmetric configuration with
only one angular momentum turned on.

Our strategy will be as follows. We shall work in five dimensions, or
equivalently, in type IIB compactified on a 5-sphere. In the absence of
R-charge, the simplest ansatz to consider is one in which only the five
dimensional metric degrees of freedom are excited. In the full type IIB
description, this is equivalent to work in a Freund--Rubin ansatz
\cite{FR}. The
general five-dimensional Kerr-AdS black holes \cite{HHT}  are then the
appropriate  configurations  to consider, since they  have the correct
asymptotics and carry the correct charges. These are families of
configurations characterised by three charges : the mass and two angular
momenta, which are in one to one correspondence with  respective three
parameters $\{m,\,a,\,b\}$. In general these configurations are not
supersymmetric.

In order to obtain supersymmetric configurations for this class of solutions,
we shall analyse particular {\it scaling limits}, for which the charges (mass,
two angular momenta) remain {\it finite}.  The finite charges, associated with
this scaled space-time saturate the BPS algebra bound. From a geometrical
point of view, the scaling limit corresponds to pushing the horizon of the
original Kerr-AdS black hole to infinity. As a result, one expects to find a
naked singularity. Nevertheless in this rescaled space time the asymptotics
corresponds to the AdS space time.

Our main result is the explicit construction of the  Killing spinors for the
 space-time  in the scaling limit with two equal, finite angular
momenta, thus proving explicitly that these are supersymmetric space-times
corresponding to the  the  one quarter BPS configuration. The solution has
 a point-like naked singularity, and asymptotes to AdS. It
corresponds to a Lorentzian Einstein-Sasaki manifold, which is in agreement
with the general statement that any five dimensional Lorentzian space-time in
a Freund--Rubin Ansatz is either locally AdS, or a Lorentzian Einstein-Sasaki
manifold or a space-time, conformal to a pp-wave \cite{felipe}.

The paper is organized in the following way.  In section \ref{sec:fived}, we
discuss the scaling limit in detail, write down the corresponding geometry,
and present the Killing spinors that support our claim. The technical details
are discussed in the appendix. In section \ref{sec:conjecture}, we comment on
the existence of a second scaling limit, giving rise to a potentially one-half BPS
configuration with a single angular momentum turned on. We comment on
the puzzles associated with this configuration if we keep the AdS
asymptote and on the emergence of a spacetime being conformal to a
pp-wave if the physical parameter scaling is done together with a rescaling of a
``lightcone'' coordinate.

\section{Scaling Limit of  Kerr--Anti-de Sitter black holes}
 \label{sec:fived}

The starting point of our analysis is the general five-dimensional
Kerr--Anti-de Sitter black hole \cite{HHT} :
\begin{multline}
  ds_5^2 = -\frac{\Delta}{\rho^2}\left[dt -
  \frac{a\sin^2\theta}{\Xi_a}\,d\phi -
  \frac{b\cos^2\theta}{\Xi_b}\,d\psi\right]^2 +
  \frac{\Delta_\theta\,\sin^2\theta}{\rho^2}\left[a\,dt -
  \frac{r^2+a^2}{\Xi_a}\,d\phi\right]^2 \\
  + \frac{\Delta_\theta\,\cos^2\theta}{\rho^2}\left[b\,dt -
  \frac{r^2+b^2}{\Xi_b}\,d\phi\right]^2 + \frac{\rho^2\,dr^2}{\Delta}
  + \frac{\rho^2\,d\theta^2}{\Delta_\theta} \\
  + \frac{(1+r^2\ell^{-2})}{r^2\rho^2}\left[
  ab\,dt - \frac{b(r^2+a^2)\sin^2\theta}{\Xi_a}\,d\phi -
  \frac{a(r^2+b^2)\cos^2\theta}{\Xi_b}\,d\psi\right]^2~,
 \label{eq:kerrmetric}
\end{multline}
where
\begin{equation}
  \begin{aligned}[m]
    \Delta &= \frac{1}{r^2}(r^2+a^2)(r^2+b^2)(1+r^2\ell^{-2})-2m~, \\
    \Delta_\theta &= 1-a^2\ell^{-2}\cos^2\theta
    -b^2\ell^{-2}\sin^2\theta~, \\
    \rho^2 &= r^2 + a^2\cos^2\theta + b^2\sin^2\theta~, \\
    \Xi_a &= 1-a^2\ell^{-2}~, \quad \Xi_b = 1-b^2\ell^{-2}~.
  \end{aligned}
 \label{eq:def1}
\end{equation}
Following the thermodynamical analysis done in \cite{GPP}, the
conserved charges carried by this configuration (energy $E$ and
angular momenta $\{J_1,\,J_2\}$) are given by
\begin{equation}
  E = \frac{\pi\,m\left(2\Xi_a + 2\Xi_b -
  \Xi_a\,\Xi_b\right)}{4\Xi_a^2\,\Xi_b^2}~, \quad
  J_1 = \frac{\pi\,m\,a}{2\Xi_a^2\,\Xi_b}~, \quad
  J_2 = \frac{\pi\,m\,b}{2\Xi_a\,\Xi_b^2}~.
 \label{eq:charges}
\end{equation}

One of the motivations for this work is the observation  that in the above
expressions there is   a non-trivial {\it scaling limit} of the parameters
$\{m,\,a,\,b\}$, for which  this configuration that keeps all charges
\eqref{eq:charges} {\it finite} and it saturate the Bogomolnyi bound. This
scaling limit corresponds to:
\begin{equation}
  a~,b\to \ell~, \quad M\equiv \frac{m}{\Xi^3}~, \text{fixed}~.
 \label{eq:abslimitd5}
\end{equation}
Specifically, we take the limit,   for which  $a$ and $b$ approach $\ell$ at
the same rate. In this case $\Xi \equiv \Xi_a=\Xi_b\to 0$. The novelty in the
above scaling limit is that we allow ourselves to scale the mass parameter $m
\to 0$ as both angular momentum parameters $a$ and $b$ reach their extremal
values $\ell$. After this scaling limit,  the physical charges of the
configuration  are {\it finite} and satisfy the relation :
\begin{equation*}
  E = \pi\,M~, \quad J_1=J_2 = \frac{\ell}{2} E~,
\end{equation*}
and thus
\begin{equation}
  E\cdot \ell = J_1 + J_2~.
 \label{eq:d5bound1}
\end{equation}
Employing the  supersymmetry algebra (with R-charges turned off), as done in
\cite{gary}, the  the eigenvalues $\{\lambda\}$ of the Bogomolnyi matrix are
given by:
\begin{equation*}
  \lambda = E \pm \frac{J_1}{\ell} \pm \frac{J_2}{\ell}~,
  \label{eq:bog}
\end{equation*}
where all signs are uncorrelated. It is now obvious that \eqref{eq:d5bound1}
indeed saturates a supersymmetry bound, and that the number of preserved
supercharges, should they exist, would be one quarter of the original ones,
since there is a single vanishing eigenvalue when the limit
\eqref{eq:abslimitd5} is considered. In the following subsection, we will
explicitly prove that our configuration \eqref{eq:abd5metric} is a one quarter
BPS one by constructing its Killing spinors, thus matching the purely
algebraic analysis mentioned above.

Concerning the nature of the metric after the scaling
\eqref{eq:abslimitd5}, it can most easily be captured by working in
a coordinate system which is non-rotating at infinity \cite{HHT}.
This is defined by
\begin{equation}
  \begin{aligned}[m]
    \Xi_a\,y^2\,\sin^2\hat{\theta} &= (r^2+a^2)\,\sin^2\theta~, \\
    \Xi_b\,y^2\,\cos^2\hat{\theta} &= (r^2+b^2)\,\cos^2\theta~, \\
    \hat{\phi} &= \phi + a\ell^{-2}\,t~, \\
    \hat{\psi} &= \psi + b\ell^{-2}t~.
  \end{aligned}
 \label{eq:cchange}
\end{equation}
Working out the change of variables, and taking the scaling limit
\eqref{eq:abslimitd5}, the final metric can be written as
\begin{multline}
  ds_5^2 = -\left(\frac{y^2}{\ell^2}+1\right)\,dt^2
  + \frac{2M}{y^2}\left(dt-\ell \sin^2\hat{\theta}
  d\hat{\Phi}-\ell \cos^2\hat{\theta}\hat{d\Psi}\right)^2 \\
  + \frac{dy^2}{1+\frac{2M\ell^2}{y^4}+\frac{y^2}{\ell^2}}
  + y^2\left(d\hat{\theta}^2+\sin^2\hat{\theta} d\hat{\Phi}^2
  +\cos^2\hat{\theta} d\hat{\Psi}^2\right)~.
 \label{eq:abd5metric}
\end{multline}

This metric has a curvature singularity at $y=0$. It is naked due to the
absence of horizons, and has no closed time-like curves. Let us notice that
from the original Kerr-AdS black hole perspective, the scaling limit
\eqref{eq:abslimitd5} is effectively pushing the horizon of the black hole to
infinity, since the mass parameter $m$, responsible for its finiteness of the
horizon,  was scaled to zero. From this perspective, it is natural to expect
that the space-time has a naked singularity. Note however, that the scaling
still allows us to asymptotically ($y\to \infty$) reach the AdS space-time, in
static coordinates.

[It turns out that the above metric can be obtained as a lorentzianisation,
i.e.  $t\to i t$ and $\ell\to -i\ell$, of specific Einstein-Sasaki metrics
obtained in \cite{Gauntlett,GauntletMartelliWaldram}. Discussions along this
direction can be found in \cite{HSY} \footnote{We thank Chris Pope for the
communication on this point.}. Our motivation, though, was primarily driven
from a purely Lorentzian point of view.]

\subsection{Supersymmetry}
 \label{sec:susy}

Since the configuration \eqref{eq:abd5metric} does not carry any R-charge, the
existence of supersymmetry can be answered by analysing the existence of
non-trivial solutions to the Killing spinor equation. The latter is the
standard Killing spinor equation for spaces with a negative cosmological
constant :
\begin{equation}
  \hat{\nabla}_\mu \eta= \left(\nabla_\mu +
  \frac{1}{2\ell}\gamma_\mu\right) \eta=0~,
 \label{eq:kspinor}
\end{equation}
where $\nabla_\mu = \partial_\mu +
\frac{1}{4}\omega_\mu{}^{ab}\gamma_{ab}$ stands for the standard
covariant derivative on spinors.

It will turn out very convenient for our purposes to analyse the set
of constraints imposed by the first two integrability conditions
associated with \eqref{eq:kspinor} :
\begin{equation}
  \begin{aligned}[m]
    [\hat{\nabla}_\mu,\,\hat{\nabla}_\nu]\eta &=& 0~, \\
    \big[\hat{\nabla}_\lambda,\,[\hat{\nabla}_\mu,\,\hat{\nabla}_\nu]\big]
    \eta &=& 0~.
  \end{aligned}
\end{equation}
These are equivalent, respectively, to \cite{vanNieuwenhuizen:1983wu}
\begin{equation}
  \begin{aligned}[m]
    C_{\mu\nu\rho\sigma}\gamma^{\rho\sigma}\eta &=& 0~, \\
    \left(\nabla_\lambda C_{\mu\nu\rho\sigma}\right)
    \gamma^{\rho\sigma}\eta + \frac{1}{\ell}
    C_{\mu\nu\lambda\rho}\gamma^\rho\eta &=& 0~,
  \end{aligned}
 \label{eq:intcond}
\end{equation}
where $C_{\mu\nu\rho\sigma}$ stands for the components of the Weyl tensor.

Out of these equations, we derive two inequivalent algebraic
projections conditions :
\begin{eqnarray}
  \left(f_1(y)\gamma_{12}-f_2(y)\gamma_{35}-f_3(y)\gamma_{24}\right)
  \eta &=& 0~, \label{eq:int1} \\
  i\gamma_{23}\eta &=& -\eta~,
  \label{eq:int2}
\end{eqnarray}
where the three scalar functions $f_i(y)$ $i=1,2,3$ are defined by
\begin{equation*}
  f_1(y) = \frac{\sqrt{y^6+\ell^2
    y^4+2M\ell^4}}{y\,\sqrt{y^4+\ell^2 y^2-2M\ell^2}}~, \quad
    f_2(y) = \frac{\ell}{y}~, \quad
    f_3(y) = \frac{y^2+\ell^2}{\sqrt{y^4+\ell^2 y^2-2M\ell^2}}~,
\end{equation*}
and we already used the fact that the five dimensional gamma matrix
$\gamma_5$ can be expressed as $i\gamma_{1234}$.

That these equations are indeed projection conditions can be trivially
realised for \eqref{eq:int2}, whereas for \eqref{eq:int1}, it follows
from the identity
\begin{equation*}
  (f_1(y))^2 + (f_2(y))^2 = (f_3(y))^2~.
\end{equation*}
Notice that the subspace of solutions of each projection condition are
orthogonal, which implies that only one quarter of all available spinors
do satisfy both equations. This orthogonality can be exposed in a more
manifest way by realising that the space of solutions to
\eqref{eq:int1} is equivalent to the space of solutions of
\begin{equation*}
  \left(f_3(y)\gamma_{14} + if_2(y)\gamma_4 \right) \eta =
  f_1(y)\eta~.
\end{equation*}
It is now clear that both $\gamma_{14}$ and $i\gamma_4$ commute with
$i\gamma_{23}$.

In the appendix \ref{sec:tech}, present  the explicit resolution of the
Killing spinor equations \eqref{eq:kspinor}. It is proved there that the
answer is given by
\begin{equation}
  \Psi = e^{-it/2\ell}\tilde{\Psi}(y)~,
 \label{eq:answer}
\end{equation}
where the spinor $\tilde{\Psi}(y)$ satisfies the differential equation
\begin{equation}
  \frac{d\tilde{\Psi}}{dy} + \frac{1}{2}
  \frac{2M\ell^2(\ell^2+3y^2)-y^4(\ell^2+y^2)}{(y^4+y^2\ell^2-2M\ell^2)\,
  \sqrt{y^6+y^4\ell^2+2M\ell^4}}\,\gamma_{14}\,\tilde{\Psi} = 0~,
\end{equation}
and fulfills both integrability conditions \eqref{eq:int1} and
\eqref{eq:int2}. Notice that this differential equation is regular
at $y=0$, where the naked singularity is.

\section{Discussion}
 \label{sec:conjecture}

In this note, we have analysed the scaling limit \eqref{eq:abslimitd5} of the
general five dimensional Kerr-AdS black hole, and we obtained a singular,  one
quarter supersymmetric space-time configuration with finite energy and two
equal, finite angular momenta. The singularity is point-like; it would be
interesting to understand whether there is any source in string theory that
could provide a physical interpretation of this singularity, analogous
to the R-charged, non-spinning BPS configuration for the superstar \cite{myers}.
Such a microscopical understanding of the  solutions, studied in this paper,
would in turn clarify whether this  singular space-time is actually a solution
of the full string theory.

Besides the scaling limit analysed previously, inspection of the
thermodynamical quantities \eqref{eq:charges} suggests the possibility of a
second, inequivalent scaling limit given by
\begin{equation}
  a\to \ell~, \quad M\equiv \frac{m}{\Xi_a^2}~,b~~ \text{fixed}~.
 \label{eq:aslimitd5}
\end{equation}
The conserved charges for such a configuration are finite and  satisfy the
following relations :
\begin{equation*}
  E = \frac{\pi}{2}\,\frac{M}{\Xi_b}~, \quad J_a = \ell\,E~, \quad
  J_b=0~.
\end{equation*}
Notice that the second angular momentum was sent to zero even though the
parameter $b$ was kept fixed and different from $\ell$. Thus, this second
scaling limit \eqref{eq:aslimitd5} formally satisfies the identity
\begin{equation}
  E\cdot \ell = J_a~.
 \label{eq:d5bound2}
\end{equation}
Employing the supersymmetry algebra of five dimensional gauged
supergravity, one derives that
\eqref{eq:d5bound2} saturates the Bogomolnyi   bound, corresponding to two
equal Bogomolnyi matrix eigenvalues \eqref{eq:bog}.  Therefore one expects
that in the scaling limit there is a configuration that preserves one half of
the supersymmetry. In order to ensure that such a  supersymmetric
configuration is indeed obtained from a scaling limit of the metric
\eqref{eq:kerrmetric}, one follows the arguments given in \cite{gary}. Namely,
the existence of such one half supersymmetric configuration would imply the
existence of two Killing vectors, constructed out of the corresponding Killing
spinors. For that purpose,  one employs the four Killing vectors of the
five-dimensional AdS space-time:
\begin{equation}
  K_{\pm\pm} = \partial_T + \ell^{-1}\left(\eta_1\partial_{\hat{\phi}} +
  \eta_2\partial_{\hat{\psi}}\right)~,
\end{equation}
where both $\{\eta_1,\,\eta_2\}$ are uncorrelated signs. [For the
five-dimensional AdS space-time these four Killing vectors have a negative
norm  everywhere, and thus give rise to maximal supersymmetry.]  We employ
these four independent Killing  vectors to analyse the  norm of these Killing
vectors for the space-time  \eqref{eq:kerrmetric} in the above scaling limit.

Using the change of coordinates \eqref{eq:cchange}, we can rewrite these
vectors in terms of the asymptotically rotating frame coordinate system, and
compute their norms in the original metric \eqref{eq:kerrmetric} while taking
the corresponding scaling limit \eqref{eq:aslimitd5}. The four Killing vectors
are :
\begin{equation}
  K_{\eta_1\eta_2} = \partial_t +
  \eta_1\ell^{-1}\left(1-\eta_1\,a\ell^{-1}\right)
  \partial_\phi +\eta_2 \ell^{-1}
  \left(1 -\eta_2\,b\ell^{-1}\right)\partial_\psi~,
 \label{eq:kvector}
\end{equation}
and their norms are given by :
\begin{eqnarray}
  g(K_{+\eta_2},\,K_{+\eta_2}) &=& -1~, \\
  g(K_{-\eta_2},\,K_{-\eta_2}) &=& -1 + \frac{8M\,\sin^4\theta}{r^2+
  \ell^2\,\cos^2\theta + b^2\sin^2\theta}~.
\end{eqnarray}
Thus, we observe that the pair of Killing vectors $K_{+\eta_2}$ are everywhere
causal, whereas the causality of the other pair $K_{-\eta_2}$ depends on the
value of $b$. It is clear that for $b=0$, the causal character of the
corresponding vectors will be flipped somewhere in space-time, whereas for
$b\neq 0$, such property depends on the physical ratio $M/b^2$. Therefore,
this analysis substantiates the potential existence of a one-half
supersymmetric solution in the above scaling limit.

Despite these algebraic facts, if one attempts to construct a finite
metric (even for the $b=0$ case) with the right AdS asymptotics, and
the right cross-terms to describe non-trivial angular momentum, one
apparently seems to require to scale $M$ to infinity, in which case
the limiting spacetime would have infinite charges.

A different possibility can arise if one changes the asymptotics of
the corresponding scaled spacetime. Let us consider the $b=0$ case in 
\eqref{eq:aslimitd5} and introduce the following ``light-cone'' coordinates:
\begin{equation}
  x^+=t-a{\hat \phi}\ , \ \ \ \  x^-=t+a{\hat \phi} \,,
 \label{eq:lcone}
\end{equation}
where ${\hat {\phi}}=\phi+a\, \ell^{-1}\, t$ is the angular coordinate
in the asymptotically static space-time. If the scaling limit 
\eqref{eq:aslimitd5}, namely: 
\begin{equation} 
  a=\ell-\ell\, \epsilon \,, \ \ \ 
  m=M(1-a^2\ell^{-2})^2\to 4\, M\, \epsilon^2\, , \ \ \ 
  \epsilon \to 0\,,
\end{equation} 
is done together with the rescaling of the $x^-$ light-cone coordinate
\begin{equation} 
  {\hat x}^{-}=\frac{x^{-}}{2\epsilon}\, , \ \ \ \epsilon \to 0,
\end{equation}
one obtains, as a consequence, that the Kerr-AdS metric
\eqref{eq:kerrmetric} becomes:
\begin{eqnarray}
  ds^2 &=& \frac{2M\ell^{-2}\sin^4\theta+\textstyle{\frac{1}{4}}
  (\cos^2\theta +r^2\ell^{-2})[-\cos^2\theta\,(1+2r^2\ell^{-2})
  +r^2\ell^{-2}]}{\cos^2\theta + r^2\ell^{-2}}\,d{ x}^+\,d{ x}^+ \\
  &-& \sin^2\theta(1+r^2\ell^{-2})\, d{ x}^+\, d{\hat x}^-\,
  +\,\frac{\cos^2\theta +r^2\ell^{-2}}{(1+r^2\ell^{-2})^2}\, dr^2\, +\,
  \frac{\cos^2\theta+r^2\ell^{-2}}{\sin^2\theta\,\ell^{-2}}\, d\theta^2\,
  +\,r^2\cos^2\theta\, d\psi^2\,.\nonumber
\end{eqnarray}
If the above metric is supersymmetric, by the general results in
\cite{felipe}, it should be conformal to a five dimensional
pp-wave. Thus, we expect the existence of a coordinate transformation 
making this fact manifest.

We would want to conclude that for Kerr AdS black holes in other dimensions
\cite{GLPP} analogous scaling limits to the ones considered here, with
all angular momenta turned on would have a straightforward
generalization. We expect solutions with finite energy and all angular
momenta equal to be singular, but asymptotic to AdS, and to preserve
some amount of supersymmetry.

\section*{Acknowledgments}
We would like to thank Ofer Aharony, Joe Minahan, Hong L\"u and Chris Pope for
useful discussions and comments.  The work is supported by the Department of
Energy under Grant No.~DE-FG02-95ER40893, by the National Science Foundation
under Grant Nos.~INT02-03585 (MC), PHY-0331728 (JS) and OISE-0443607 (JS) as
well as by the Fay R. and Eugene L. Langberg Chair  (MC).

\newpage

\appendix

\section{Analysis of Killing spinor equations}
 \label{sec:tech}

In this appendix, we shall provide the explicit details giving rise to
the Killing spinors associated to the scaling solution
\eqref{eq:abd5metric}. Thus, we need first to compute the spin
connection associated with this metric. Working with the vielbein
basis :
\begin{eqnarray}
  e^1 &=& h_2\,dt + 2M\frac{\ell}{y^2}h_2^{-1}\left(\sin^2\theta\,d\phi
  + \cos^2\theta\,d\psi\right)~, \nonumber \\
  e^2 &=& \frac{y}{2}\sin(\phi+\psi)\,\sin
  2\theta\,\left(d\phi-d\phi\right)
  + y\cos(\phi+\psi)\,d\theta~, \nonumber \\
  e^3 &=& \frac{y}{2}\cos(\phi+\psi)\,\sin 2\theta\,
  \left(d\psi -d\phi\right) +
  - y\sin(\phi+\psi)\,d\theta~, \nonumber \\
  e^4 &=& h_3\left(\sin^2\theta\,d\phi + \cos^2\theta\,d\psi\right)~,
  \nonumber \\
  e^5 &=& h_1\,dy~,
 \label{eq:vield5}
\end{eqnarray}
where we introduced the set of functions
\begin{multline*}
  h_2(y) = \left(1-2\frac{M}{y^2}+\frac{y^2}{\ell^2}\right)^{1/2}~, \,\,
  h_3(y) = \left(\frac{y^6+2M\ell^4 +
  \ell^2y^4}{y^4+\ell^2y^2-2M\ell^2}\right)^{1/2}~, \\
  h_1(y)\cdot h_2(y)\cdot h_3(y) = y~,
\end{multline*}
the spin connection $\omega^a\,_b$ solving the algebraic equation
\begin{equation*}
  de^a + \omega^a\,_b\wedge e^b = 0~,
\end{equation*}
is given by
\begin{eqnarray}
  \omega^1\,_2 &=& -\frac{2M\ell}{y^4\,h_2}\, e^3~, \,\,
  \omega^1\,_3 = \frac{2M\ell}{y^4\,h_2}\, e^2~, \nonumber \\
  \omega^1\,_4 &=& \frac{2M\ell}{y^4}\,\left[1+
  y\frac{h_2'}{h_2}\right]\,e^5=F(y)\,e^5 \nonumber \\
  \omega^1\,_5 &=& \frac{h_2'}{h_2\,h_1}\,e^1 - F(y)\,e^4~,
  \nonumber \\
  \omega^2\,_3 &=& -\frac{2M\ell}{y^4\,h_2}\,e^1 +
  \left(\frac{h_3}{y^2}-\frac{2}{h_3}\right)\,e^4~, \nonumber \\
  \omega^2\,_4 &=& \frac{h_3}{y^2}\,e^3~, \,\,
  \omega^3\,_4 = -\frac{h_3}{y^2}\,e^2~, \nonumber \\
  \omega^2\,_5 &=& \frac{1}{y\,h_1}\,e^2~, \,\,
  \omega^3\,_5 = \frac{1}{y\,h_1}\,e^3~, \nonumber \\
  \omega^4\,_5 &=& \frac{h_3'}{h_3\,h_1}\,e^4 + F(y)\,e^1~,
\end{eqnarray}
where the function $F(y)$ appearing in the equations was defined in
the second line above and all primes indicate derivative with respect
to $y$.

We are now at a position to study the different components of the
Killing spinor equation
\begin{equation*}
  \nabla_\mu\Psi = -\frac{1}{2\ell}\gamma_\mu\,\Psi~.
\end{equation*}
First, consider the time component $(\mu=t)$ equation :
\begin{equation}
  \partial_t\Psi +\frac{1}{2}\left(h_2'h_1^{-1}\,\gamma_{15} +
  F(y)h_2\,\gamma_{45}- 2M\frac{\ell}{y^4}\,\gamma_{23}\right)\,\Psi =
  -\frac{1}{2\ell}h_2\,\gamma_1\,\Psi~.
 \label{eq:kt}
\end{equation}
This is an equation that can be immediately integrated in terms of an
exponential function, due to the Killing vector nature of
$\partial_t$. Instead of proceeding in that way, we shall study the
matrix acting on the Killing spinor whenever the two integrability
conditions \eqref{eq:int1} and \eqref{eq:int2} are satisfied. It can
be shown that
\begin{equation*}
  \left(h_2'h_1^{-1}\,\gamma_{15} +
  F(y)h_2\,\gamma_{45}- 2M\frac{\ell}{y^4}\,\gamma_{23}
  +\frac{1}{\ell}h_2\,\gamma_1\right)\,\Psi = \frac{i}{\ell}\,\Psi~.
\end{equation*}
Thus, whenever the integrability conditions are satisfied, we can
integrate \eqref{eq:kt} :
\begin{equation}
  \Psi = e^{-it/2\ell}\tilde{\Psi}~.
 \label{eq:ans1}
\end{equation}

Let us consider next the $\mu=\theta$ component :
\begin{equation}
  \partial_\theta\Psi
  + \left(M\frac{\ell}{y^3\,h_2}\,\gamma_{13}
  -\frac{h_3}{2y}\,\gamma_{34} + \frac{2}{h_1}\,\gamma_{25}
  +\frac{y}{2\ell}\,\gamma_2\right)\,e^{-(\phi+\psi)\gamma_{23}}\,
  \Psi =0~.
 \label{eq:ktheta}
\end{equation}
It is amusing to realise that whenever both integrability conditions
\eqref{eq:int1} and \eqref{eq:int2} are satisfied, the following
identity holds :
\begin{equation}
  \left(M\frac{\ell}{y^3\,h_2}\,\gamma_{13}
  -\frac{h_3}{2y}\,\gamma_{34} + \frac{2}{h_1}\,\gamma_{25}
  +\frac{y}{2\ell}\,\gamma_2\right)\,\Psi = 0~.
 \label{eq:iden1}
\end{equation}
Therefore, we conclude the Killing spinor $\Psi$ is independent of the
$\theta$ angular variable $(\partial_\theta\Psi=0)$.

Let us jointly consider the two components involving the two angular
variables $\{\phi,\,\psi\}$. For the first one, we have
\begin{multline}
  \partial_\phi\Psi +\frac{1}{2}\sin
  2\theta\left[\frac{M\,\ell}{y^3\,h_2}\,\gamma_{12} - \frac{h_3}{2y}\,
  \gamma_{24}-\frac{1}{2h_1}\,\gamma_{35}-\frac{y}{2\ell}\,\gamma_3\right]
  \,e^{-(\phi+\psi)\gamma_{23}}\,\Psi \\
  +\frac{1}{2}\sin^2\theta\left[\left(\frac{2M\,\ell}{y^3}\,\frac{h_3}{h_2}\,h_2'
  -h_3\,F(y)\right)\gamma_{15} +
  \left(\left(\frac{h_3}{y}\right)^2-2-\left(\frac{2M\,\ell}{y^3\,h_2}\right)^2\right)
  \gamma_{23} \right. \\
  \left. +\left(\frac{2M\,\ell}{y^2\,h_2}\,F(y)+\frac{h_3'}{h_1}\right)\gamma_{45}
  + \frac{2M}{y^2\,h_2}\,\gamma_1 + \frac{h_3}{\ell}\,\gamma_4\right]\,\Psi
 \label{eq:kphi}
\end{multline}
whereas for the second one:
\begin{multline}
  \partial_\psi\Psi -\frac{1}{2}\sin
  2\theta\left[\frac{M\,\ell}{y^3\,h_2}\,\gamma_{12} - \frac{h_3}{2y}\,
  \gamma_{24}-\frac{1}{2h_1}\,\gamma_{35}-\frac{y}{2\ell}\,\gamma_3\right]
  \,e^{-(\phi+\psi)\gamma_{23}}\,\Psi \\
  +\frac{1}{2}\cos^2\theta\left[\left(\frac{2M\,\ell}{y^3}\,\frac{h_3}{h_2}\,h_2'
  -h_3\,F(y)\right)\gamma_{15} +
  \left(\left(\frac{h_3}{y}\right)^2-2-\left(\frac{2M\,\ell}{y^3\,h_2}\right)^2\right)
  \gamma_{23} \right. \\
  \left. +\left(\frac{2M\,\ell}{y^2\,h_2}\,F(y)+\frac{h_3'}{h_1}\right)\gamma_{45}
  + \frac{2M}{y^2\,h_2}\,\gamma_1 + \frac{h_3}{\ell}\,\gamma_4\right]\,\Psi~.
 \label{eq:kpsi}
\end{multline}

It is important to realise that due to the identity \eqref{eq:iden1},
the matrix multiplying the $\sin 2\theta$ terms in both equations
\eqref{eq:kphi} and \eqref{eq:kpsi} vanishes identically. We are thus
only left to evaluate the matrix multiplying both $\sin^2\theta$ and
$\cos^2\theta$. Such matrix can be written, after using the second
integrability condition \eqref{eq:int2} as
\begin{equation}
  A(y)\gamma_4 + i\left(\left(\frac{h_3}{y}\right)^2-2-
  \left(\frac{2M\,\ell}{y^3\,h_2}\right)^2\right) + B(y)\gamma_1~,
 \label{eq:step}
\end{equation}
where the following definitions and identities hold :
\begin{eqnarray*}
  A(y) &=& \frac{2M\ell}{y^3}\frac{h_2'}{h_2\,h_3}-h_3\,F(y) +
  \frac{h_3}{\ell} = \frac{y^4-2M\ell^2}{\ell\,y^3}\,f_1(y)~, \\
  B(y) &=& \frac{2M\ell}{y^2\,h_2}\left(F(y)+\frac{1}{\ell}\right)
  + \frac{h_3'}{h_1} = \frac{y^4-2M\ell^2}{\ell\,y^3}\,f_3(y)~.
\end{eqnarray*}
Since the second integrability condition \eqref{eq:int1} is equivalent
to
\begin{equation*}
  \left(f_1(y)\gamma_4 + f_3(y)\gamma_1\right)\Psi =
  i\frac{\ell}{y}\Psi~,
\end{equation*}
the matrix $A(y)\gamma_4 + B(y)\gamma_1$, when acting on the Killing
spinor $\Psi$ satisfies the identity
\begin{equation*}
  \left(A(y)\gamma_4 + B(y)\gamma_1\right)\Psi = i
  \frac{y^4-2M\ell^2}{y^4}\Psi~.
\end{equation*}
It turns out that the above term is minus the second term in
\eqref{eq:step}. Therefore, the sum of all matrices acting on Killing
spinors appearing in \eqref{eq:kphi} and \eqref{eq:kpsi} vanish
whenever both integrability conditions \eqref{eq:int1} and
\eqref{eq:int2} are satisfied. We conclude the Killing spinor is also
independent of the angular variables $\{\phi,\,\psi\}$.

Finally, let us focus on the radial $(\mu=y)$ component equation :
\begin{equation}
  \partial_y\Psi + \frac{h_1}{2}\left(F(y)\,\gamma_{14}
  + \frac{1}{\ell}\,\gamma_5\right)\Psi = 0~.
 \label{eq:ky}
\end{equation}
When using the partial integration \eqref{eq:ans1} and the second
integrability condition \eqref{eq:int2}, we learn that
\begin{equation*}
  \frac{d\tilde{\Psi}}{dy} + \frac{h_1}{2}\left(F(y)-
  \frac{1}{\ell}\right)\gamma_{14}\tilde{\Psi} = 0~.
\end{equation*}
The latter is a first order differential equation that can always be
integrated. Evaluating both functions $F(y)$ and $h_1(y)$, its
explicit expression is :
\begin{equation}
  \frac{d\tilde{\Psi}}{dy} + \frac{1}{2}
  \frac{2M\ell^2(\ell^2+3y^2)-y^4(\ell^2+y^2)}{(y^4+y^2\ell^2-2M\ell^2)\,
  \sqrt{y^6+y^4\ell^2+2M\ell^4}}\,\gamma_{14}\,\tilde{\Psi} = 0~,
\end{equation}

To sum up, we have proved, by explicit construction, the existence of
non-trivial Killing spinors for the background \eqref{eq:abd5metric}.
For the reasons discussed in the main text, we conclude that such a
background preserves one quarter of the supersymmetry.

\bibliographystyle{utphys}
\bibliography{biblio1}

\end{document}